\documentstyle[twoside,fleqn,espcrc2,psfig]{article}

\newcommand{\NP}{{\em Nucl.\ Phys.\ }}
\newcommand{\PL}{{\em Phys.\ Lett.\ }}
\newcommand{\PR}{{\em Phys.\ Rev.\ }}
\newcommand{\PRP}{{\em Phys.\ Rep.\ }}

\newcommand{\MPL}{{\em Mod.\ Phys.\ Lett.\ }}
\newcommand{\PRL}{{\em Phys.\ Rev.\ Lett.\ }}

\newcommand{\um}[1]{\"{#1}}
\newcommand{\tr}{{\rm Tr}}

\newcommand{\fslash}{F\!\!\!\!/\ }
\newcommand{\inn}{\!\cdot\!}
\newcommand{\norm}[1]{\raise.3ex\hbox{:}#1\raise.3ex\hbox{:}}
\def\binomial#1#2{\left( #1 \atop #2 \right)}

\title{Scattering of Strings from D-branes}

\author{Akikazu Hashimoto and Igor R. Klebanov%
\address{ Joseph Henry Laboratories \\ 
          Princeton University \\ 
          Princeton, New Jersey 08544}
}

\begin{document}

\begin{abstract}
We review a number of perturbative calculations describing the
interactions of D-branes with massless elementary string states. The
form factors for the scattering of closed strings off D-branes are
closely related to the Veneziano amplitude.  They show that, in
interactions with strings, D-branes acquire many of their physical
features: the effective size of D-branes is of order the string scale
and expands with the energy of the probe, while the fixed angle
scattering amplitudes fall off exponentially.  We also calculate the
leading process responsible for the absorption of closed strings: the
amplitude for a closed string to turn into a pair of open strings
attached to the D-brane. The inverse of this process describes the
Hawking radiation by an excited D-brane.
\end{abstract}

\maketitle


Recent developments in the study of string theory at a
non-perturbative level have shown that string theory is not just a
theory of strings, but it also contains extended object of higher
dimensionality known as the $p$-branes
\cite{filq,sen,schwarz,Schwarz2,duff,chpt,witten,strominger,jhas}.
Since they are exchanged with the elementary strings under
non-perturbative duality symmetries, the $p$-brane degrees of freedom
are just as indispensable for the overall consistency of the theory.
The $p$-branes have been known for some time as soliton solutions to
the low-energy effective action of string theory
\cite{dh,mdjl,chs,hs,Duff:1995}.  In this description it was
difficult, however, to investigate their `stringy' properties.  The
situation has changed dramatically following Polchinski's recent
observation\cite{polchinski} that the $p$-branes carrying
Ramond-Ramond (R-R) charges admit a remarkably simple world sheet
description in terms of open strings with Dirichlet boundary
conditions \cite{dlp,Green:1991,mgpw,Li:1994,Gutperle:1995}.  The
essential idea is the following.  Even in type II theories one
introduces world sheets with boundaries, imposing the Neumann boundary
conditions on on coordinates $X^m$ for $0 \le m \le p$ and the
Dirichlet boundary conditions on coordinates $X^M$ for $p+1 \le M \le
9$.\footnote{It will be necessary to distinguish between components of
space-time vectors that are parallel or transverse to the $p$-brane.
In these notes we will use three different types of indices,
$(m,M,\mu)$, and follow the convention that $0 \le m \le p$, $p+1 \le
M \le 9$, and $0 \le \mu \le 9$.}  Thus, the end-points of the open
strings are free to move along a $p+1$-dimensional hypersurface
defined by $X^M=a^M$ for $p+1 \le M \le 9$. This hypersurface is to
be thought of as the world volume of a $p$-brane, which has been named
a Dirichlet brane (or D-brane). Since the Dirichlet boundary
conditions preserve conformal invariance, the D-branes are exact
solutions of the string tree level equations of motion. Therefore,
this simple world sheet formulation includes the physics of all the
string modes, i.e. the D-branes are exact embeddings of the R-R charged
low-energy soliton solutions into string theory.

The D-branes are dynamical objects, whose transverse positions are
specified by collective coordinates $a^M$, and whose fluctuations are
described by the excitations of the open strings attached to them.
Their masses scale as $1/g$ which suggests that they are responsible
for the non-perturbative effects of strength $e^{1/g}$ generally
present in string physics \cite{Shenker:1990,jp2}.  The fact that the
D-branes carry R-R charges implies that they are exchanged with the
elementary strings under duality transformations which exchange the
R-R gauge fields with the NS-NS gauge fields.  Indeed, many
non-perturbative phenomena find a simple explanation in the language
of D-branes. Interested readers are referred to
\cite{joeReview,CliffReview,joeReview2} for reviews and lists of
references.

The D-branes have also found an important application 
to the physics of black holes.  
Indeed, a variety of black hole solutions to low energy
supergravity have an exact stringy description in terms of
intersecting D-branes. Their excitations are the open strings attached to
the D-branes, which allows for the counting of
the black hole entropy 
\cite{Strominger:1996,cm,ghas,HMS96,HLM96,gkp,JKM96,BLMPSV96,IgorTseytlin,VijayFinn}
Furthermore, these methods give a new insight into
the Hawking radiation rate of near-extremal
black holes
\cite{cm,DMW96,DasMathur96-2,DasMathur96-3,Maldacena:1996,SteveIgor}.

One of the main advantages of D-branes is that their dynamics admits a
perturbative description in the weak coupling limit.  The leading
perturbative amplitudes can be computed by evaluating correlation
functions on world sheets with one hole (a disk) and two holes (an
annulus).  Amplitudes of this kind have been computed by several
authors
\cite{KT,GHKM,gm,decay,hashimoto96,hashimoto96a,Barbon,Balasubramanian:1996,%
bachas,Lifschytz:1996a}.  In these notes, we review the simplest
calculations relevant to the perturbative dynamics of D-branes and
closed strings: those carried out on a disk.  These amplitudes capture
the physics of a single D-brane interacting with closed strings.  A
useful analogy to keep in mind is that of the fixed target scattering
experiments designed to probe the structure of the atoms or nuclei.
In this spirit, we learn something about the structure of D-branes by
performing elastic and inelastic scattering experiments using a beam
of closed strings.

\begin{figure}[tb]
\centerline{A:\qquad\parbox{\hsize}{\psfig{figure=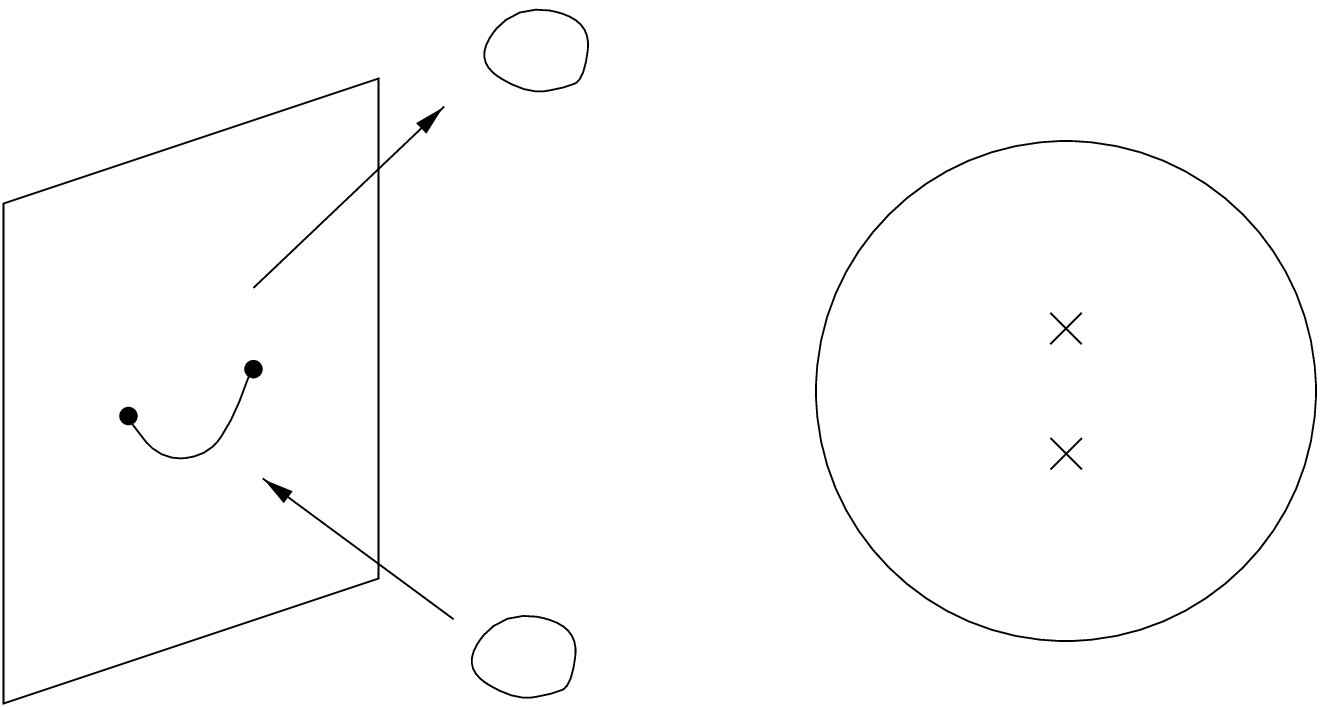,width=\hsize}}}
\centerline{B:\qquad\parbox{\hsize}{\psfig{figure=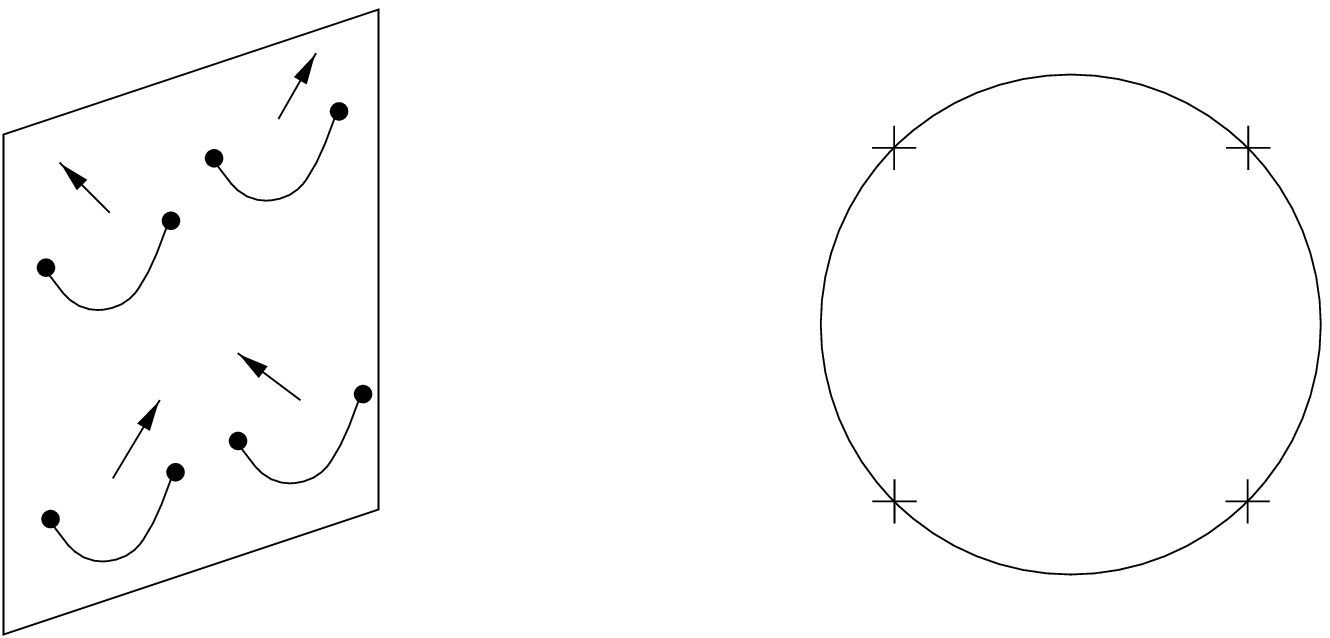,width=\hsize}}}
\centerline{C:\qquad\parbox{\hsize}{\psfig{figure=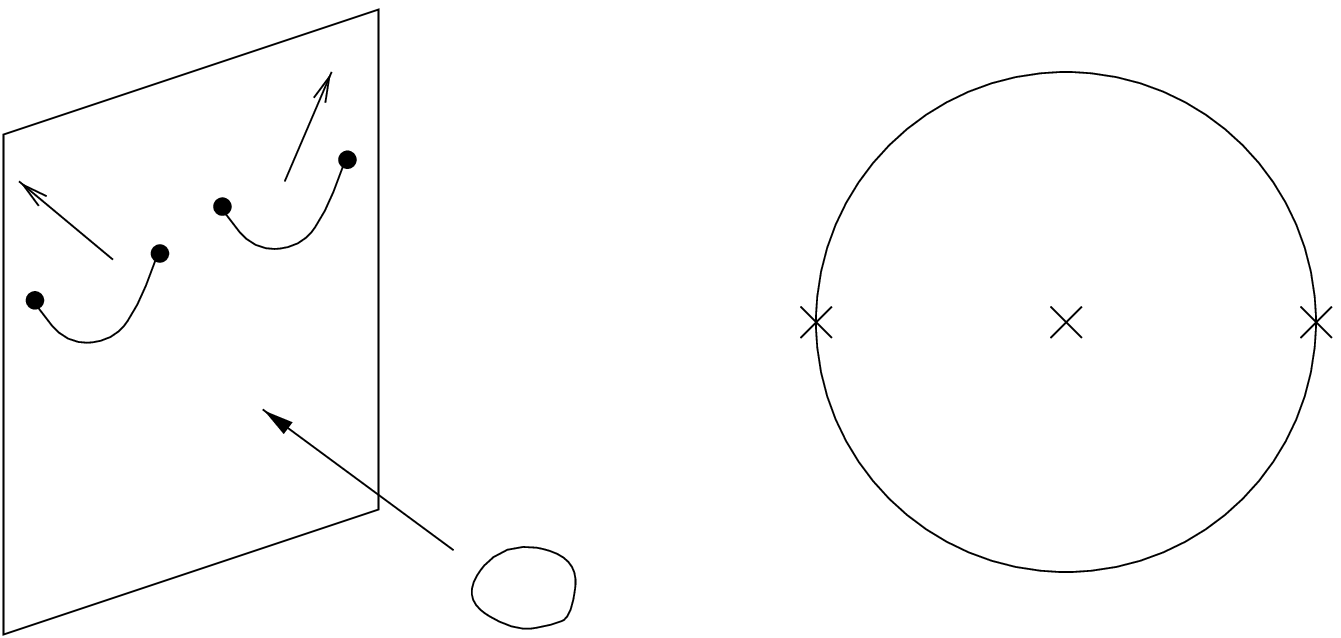,width=\hsize}}}
\caption{Schematic illustration and world sheet diagrams of the
scattering processes involving a 
single D-brane and several string excitations,
to leading order in $g$. }
\label{figa}
\end{figure}

To be more specific, we consider type II theories in a background of a
single D-brane and calculate the amplitudes illustrated in figure
\ref{figa}.  The first process corresponds to an elastic scattering of
a closed string off of a D-brane.  One can infer the size of these
D-branes from the measurement of the scattering form factors.  The
second process corresponds to a change in the internal states of the
D-branes.  The third diagram describes an inelastic process where a
closed string is absorbed by a D-brane, exciting its internal state by
creating a pair of open strings. The reverse process corresponds to
spontaneous emission by excited D-branes.  In light of the recent
realization that D-branes can be used to describe certain types of
black holes \cite{Strominger:1996,cm,ghas,HMS96,HLM96,%
gkp,JKM96,BLMPSV96,IgorTseytlin,VijayFinn}, these processes model the
classical absorption and Hawking radiation
\cite{cm,DMW96,DasMathur96-2,DasMathur96-3,Maldacena:1996,SteveIgor}.
It will turn out that, at the formal level of evaluating correlation
functions, these three types of amplitudes are closely related. Once
one is computed, the result can be easily adapted to find the other
two.

The characteristic length scale of the D-brane form factors that we
will find from our perturbative analysis is the string scale
$\sqrt{\alpha'}$.\footnote{The D-instanton is a special case
exhibiting a point-like structure.}  This is hardly a surprise given
the manifestly stringy formulation of these amplitudes.  The effective
thickness of order $\sqrt{\alpha'}$ may be visualized as a `halo' of
open strings attached to the D-branes.  The main conclusion is that
the effective size of the D-branes, as seen by the elementary strings,
is a quantity of order string length which increases with the energy
of the probe, thus exhibiting the Regge behavior \cite{regge}.

One could ask a somewhat different question: what is the
effective size of D-branes as measured by other 
D-branes?  This question was
studied in 
\cite{bachas,Lifschytz:1996a,%
Douglas:1996,Johnson:1996,DanFerrSund,PouliotKabat,DKPS96} and
length-scales shorter than the string length \cite{shenker} 
were found in cases where the coupling is weak, and the
heavy D-branes move very slowly.
For example, the scattering of two zero-branes reveals a non-perturbative
length scale $\sim g^{1/3} \sqrt{\alpha'}$, which is nothing but the
Planck length in M-theory.
Thus, using D-branes as probes reveals new aspects of their substructure,
but the necessary methods are non-perturbative and will not be covered
here.

The organization of these notes is as follows.  In section \ref{nsns}
we consider the scattering of massless NS-NS bosons illustrated in
figure \ref{figa}.A.  The Regge behavior at string scale will emerge
from this calculation.  In section \ref{rr}, we describe the extension
to massless R-R bosons.  In section \ref{pbrane}, we compare the
scattering amplitudes computed in sections \ref{nsns} and \ref{rr}
with the scattering off solitons solutions in low energy supergravity
theory.  The Dirichlet $p$-branes and the $p$-brane solutions give
rise to identical scattering at large impact parameter, confirming
that the Dirichlet $p$-branes are indeed the stringy realizations of
the $p$-branes of the low-energy supergravity.  In section
\ref{openclosed}, we describe how the type I four-point amplitude is
related to the processes illustrated in \ref{figa}.A and
\ref{figa}.B. Finally, in section \ref{decay}, we describe how the
relation between \ref{figa}.A and \ref{figa}.B can be extended to
compute the amplitudes of type \ref{figa}.C with ease. We conclude in
section \ref{conclusion}. Most of the material reviewed in these notes
has appeared in earlier papers \cite{KT,GHKM,gm,decay}.

\section{Elastic scattering of massless NS-NS bosons from D-branes}
\label{nsns}

In this section, we review the elastic scattering of a massless
NS-NS boson off a single D-brane \cite{KT}.  We will be using the
modern covariant formulation defined in terms of conformal field
theories on the world sheet \cite{FMS}.  Nice reviews of conformal
field theory approach to string theory can be found in
\cite{LustTheisen,Polchinski:1994,Dixon:1989}.

Consider a type II theory in flat ten-dimensional spacetime 
and imagine
a $p$-brane spanning the $X_1 \times X_2 \times \ldots X_p$ plane.
The presence of the $p$-brane breaks the $SO(1,9)$ Lorentz symmetry to
$SO(1,p) \times SO(9-p)$. The $p$-brane's mass scales as
$1/g$.  Therefore, to leading order in $g$, 
it is infinitely heavy and does not recoil:  it is capable of absorbing
an arbitrary amount of momentum in the $X_{p+1} \ldots X_{9}$ directions
without a change in its energy.

Now imagine an incoming and outgoing massless NS-NS bosons with
momenta $p_1$ and $p_2$, and polarizations $\varepsilon_1$ and
$\varepsilon_2$ respectively.  As we saw above, only the components of
momentum parallel to the brane are conserved.  Let us define two
kinematic quantities invariant under $SO(1,p)\times SO(9-p)$,
\begin{eqnarray}
s &=& 2 p_{1\, \parallel}^2 = 2 p_{2\,\parallel}^2 \nonumber \\
t &=& p_1 \inn p_2\ , \nonumber 
\end{eqnarray}
and compute the amplitude for a massless
NS-NS state with momentum $p_1$ to
scatter into a massless NS-NS state with momentum $p_2$. 

In the modern covariant formulation, scattering amplitudes are
computed by evaluating the correlation function of vertex operators
corresponding to the asymptotic states in the scattering process.  For
the process of the type illustrated in figure \ref{figa}.A, the
correlation function takes the form
\begin{equation} 
A =  \int \frac{d^2\!z_1\, d^2\!z_2\ }{V_{CKG}} \langle\, V_1(z_1,\bar{z}_1)
\ V_2(z_2,\bar{z}_2)\,\rangle.
\label{corr}
\end{equation}
The factor of $V_{CKG}$ accounts for the volume of the conformal
Killing group which is a residual gauge symmetry that survives after
choosing the conformal gauge.

The form of the vertex operators is constrained by the superconformal
invariance of the correlation function.  For massless NS-NS states,
they are given by
\begin{eqnarray}
V_1(z_1,\bar{z}_1)&=&\varepsilon_{1\mu\nu}\,\norm{V_{-1}^\mu(p_1,z_1)}
\ \norm{\bar{V}_{-1}^\nu(p_1,\bar{z}_1)}
\nonumber\\
V_2(z_2,\bar{z}_2)&=&\varepsilon_{2\mu\nu}\,\norm{V_{0}^\mu(p_2,z_2)}
\ \norm{\bar{V}_{0}^\nu(p_2,\bar{z}_2)}\ \ ,
\nonumber
\end{eqnarray}
with
\parbox{\hsize}{
\begin{eqnarray}
\lefteqn{V_{-1}^\mu(p_1,z_1)} \nonumber \\
&=&e^{-\phi(z_1)}\,\psi^{\mu}(z_1)\,e^{ip_1\cdot X(z_1)}
\nonumber\\
\lefteqn{V_0^\mu(p_2,z_2)} \nonumber \\
&=&\left(\partial X^\mu(z_2)+ip_2\inn \psi(z_2)\psi^{\mu}(z_2)
\right)\,e^{ip_2\cdot X(z_2)}\ \ . \nonumber 
\end{eqnarray}
}
The subscripts $\{0,-1\}$ denote the superghost charge carried by the
vertex operators.  The total amount of superghost charge on a disk is
required to be $-2$ \cite{FMS,KLS88}. This requirement is a
consequence of the superdiffeomorphism invariance on the string world
sheet.

Now that we have all the basic ingredients laid out, all that remains
to be done is to evaluate the correlation function (\ref{corr}),
keeping in mind that Neumann boundary conditions are imposed on the
directions parallel to the brane, and Dirichlet boundary conditions on
the directions transverse to the brane.  These correlation functions
are simple (but quite tedious in practice) to compute because all the
fields on the world sheet are free.

Although the original calculations of these amplitudes were performed
on a disk \cite{KT}, it turns out to be simpler to conformally
map the disk onto a half plane where both the boundary conditions and
the Greens functions take on a simple form.  Let us take our world
sheet to be the upper half plane $\cal{H}^+$.  The boundary
conditions imposed on the real axis are
\begin{eqnarray}
X(z) & = & \bar{X}(\bar{z}) \nonumber \\
\psi(z) & = & \bar{\psi}(\bar{z}) \nonumber
\end{eqnarray}
for the Neumann case and
\begin{eqnarray}
X(z) & = & -\bar{X}(\bar{z}) \nonumber \\
\psi(z) & = & -\bar{\psi}(\bar{z}) \nonumber
\end{eqnarray}
for the Dirichlet case.
These boundary conditions mix the holomorphic and the antiholomorphic
components of the world sheet fields.  Their correlation functions are
given by
\begin{eqnarray}
\langle X(z) X(w) \rangle & = & \ln(z-w) \nonumber\\
\langle X(z) \bar{X} (\bar{w}) \rangle & =  & \ln(z-\bar{w}) \nonumber  \\
\langle \psi(z) \psi(w) \rangle & = & 1/(z-w) \nonumber\\
\langle \psi(z) \bar{\psi} (\bar{w}) \rangle & =  & 1/(z-\bar{w})
\label{ncorr}
\end{eqnarray}
for a Neumann boundary and
\begin{eqnarray}
\langle X(z) X(w) \rangle & = & \ln(z-w) \nonumber\\
\langle X(z) \bar{X} (\bar{w}) \rangle & =  & -\ln(z-\bar{w}) \nonumber  \\
\langle \psi(z) \psi(w) \rangle & = & 1/(z-w) \nonumber \\
\langle \psi(z) \bar{\psi} (\bar{w}) \rangle & =  & -1/(z-\bar{w})
\label{dcorr}
\end{eqnarray}
for a Dirichlet boundary.
Now we introduce a notational device, often referred to in the
literature as the ``doubling trick.''  The fields $X(z)$ and $\psi(z)$
are originally defined only on the half plain ${\cal H}^+$. Let us imagine
extending the definition of these fields to the full plane using
$\bar{X}(\bar{z})$ and $\bar{\psi}(\bar{z})$ in a following way:
$$
X(z) = \left\{
\begin{array}{cc}
X(z) & z \in {\cal H}^+ \\
\pm\bar{X}(z) & z \in {\cal H}^- 
\end{array} \right.
$$
$$
\psi(z) = \left\{
\begin{array}{cc}
\psi(z) & z \in {\cal H}^+ \\
\pm\bar{\psi}(z) & z \in {\cal H}^- 
\end{array} \right.
$$
The choice of sign depends on the boundary condition: plus for Neumann
and minus for Dirichlet.  Now, if we think of $\bar{z}$ and $\bar{w}$
as being in ${\cal H}^-$, correlation functions 
(\ref{ncorr}) and (\ref{dcorr}) are correctly given by those of the
holomorphic fields on the entire complex plane,
\begin{eqnarray}
\langle X(z) X(w) \rangle   & = & \ln(z-w) \nonumber \\
\langle \psi(z) \psi(w) \rangle   & = & 1/(z-w) \nonumber 
\end{eqnarray}
To summarize, when
considering scattering off $p$-branes, one can replace
$\bar{X}^\mu(\bar{z})$ by ${D^\mu}_\nu X^\nu(z)$, and analogously
for the fermions,
where
$${D^\mu}_\nu = \left[
\begin{array}{cccccc}
1 &      & &  &      & \\
  &\ddots& &  &      & \\
  &      &1&  &      & \\
  &      & &-1&      & \\
  &      & &  &\ddots& \\
  &      & &  &      &-1 \\
\end{array}
\right].
$$
Now the amplitude takes
the form
\begin{eqnarray}
\lefteqn{A = \int \frac{dz_1 d\bar{z}_1 dz_2 d\bar{z}_2}{V_{CKG}}
\varepsilon_{1\mu\lambda}D^\lambda{}_\nu\, 
\varepsilon_{2\sigma\eta}D^\eta{}_\kappa}\nonumber \\
\lefteqn{\ \langle
V_{-1}^\nu(D p_1,\bar{z}_1)
V_{-1}^\mu(p_1,z_1)
V_{0}^\sigma(p_2,z_2)
V_{0}^\kappa(D p_2,\bar{z}_2)
 \rangle} \nonumber 
\end{eqnarray}
Traditionally, the conformal Killing volume in the
denominator is cancelled by fixing the positions of 
the vertex operators on
the world sheet and inserting a Koba-Nielson factor. Readers are
referred to \cite{FMS,LustTheisen,KLS88} for details.  
With a convenient
choice of the vertex operator positions, 
$$\{ \bar{z}_1, z_1, z_2, \bar{z}_2 \} = \{-iy, iy, i, -i\}$$
the amplitude reduces to
\begin{eqnarray}
\lefteqn{A = \int_0^1 \, dy\, (1-y^2)\,
\varepsilon_{1\mu\lambda}D^\lambda{}_\nu\, 
\varepsilon_{2\sigma\eta}D^\eta{}_\kappa}\nonumber \\
\lefteqn{\langle
V_{-1}^\nu(D p_1,-iy)
V_{-1}^\mu(p_1,iy)
V_{0}^\sigma(p_2,i)
V_{0}^\kappa(D p_2,-i)
 \rangle} \nonumber \\ \
\label{2camp}
\end{eqnarray}
Evaluating the necessary correlation function
we find
\begin{eqnarray}
A &=& \int_0^1 dy 
\left[ \frac{4y}{(1+y)^2} \right]^s
\left[ \frac{(1-y)^2}{(1+y)^2} \right]^t \nonumber \\
&& \qquad \times
\left[ \frac{1}{1-y^2} a_1 - \frac{1-y}{4 y (1+y)} a_2 \right]
\nonumber
\end{eqnarray}
where $a_1$ and $a_2$ contain the dependence on 
the polarizations and
momenta, but are independent of $y$.  To do the $y$-integral, it is
convenient to perform a change of variable,
$$ y = \frac{1-\sqrt{x}}{1+\sqrt{x}},$$
which brings it to the form
$$
A = \int_0^1 dx\ 
(1-x)^s x^t
(- a_1 x^{-1} + a_2 (1-x)^{-1}).
$$
This is a well known integral representation of the
Euler Beta function
\begin{equation}
A = \frac{\Gamma(t) \Gamma(s)}{\Gamma(1+s+t)} (s a_1 - t a_2)
\label{beta}
\end{equation}
For completeness, we list the explict expressions for $a_1$ and $a_2$ below:
\begin{eqnarray}
a_1 & =&
{\rm Tr}(\varepsilon_1\inn D)\,p_1\inn \varepsilon_2 \inn p_1 
-p_1\inn\varepsilon_2\inn D\inn\varepsilon_1\inn p_2 \nonumber \\
&& - p_1\inn\varepsilon_2\inn\varepsilon_1^T \inn D\inn p_1
 -p_1\inn\varepsilon_2^T \inn \varepsilon_1 \inn D \inn p_1  \nonumber \\
&&- p_1\inn\varepsilon_2\inn\varepsilon_1^T \inn p_2 +
\frac{s}{2}\,{\rm Tr}(\varepsilon_1\inn\varepsilon_2^T) \nonumber \\
&&+\Big\{1\longleftrightarrow 2\Big\}
\nonumber \\
a_2 & = &
{\rm Tr}(\varepsilon_1\inn D)\,
(p_1\inn\varepsilon_2\inn D\inn p_2 
+ p_2\inn D\inn\varepsilon_2\inn p_1  \nonumber \\
&& +p_2\inn D\inn\varepsilon_2\inn D\inn p_2)
+p_1\inn D\inn\varepsilon_1\inn D\inn\varepsilon_2\inn D\inn p_2 
\nonumber \\
&&-p_2\inn D\inn\varepsilon_2\inn\varepsilon_1^T\inn D\inn p_1 
+\frac{s}{2}\,{\rm Tr}(\varepsilon_1\inn D\inn \varepsilon_2\inn D)\nonumber \\
&&-\frac{s}{2}\,{\rm Tr}(\varepsilon_1\inn\varepsilon_2^T) \nonumber \\
&&-\frac{s+t}{2}{\rm Tr}(\varepsilon_1\inn D) {\rm Tr}(\varepsilon_2\inn D)
\nonumber \\
&&+\Big\{1\longleftrightarrow 2 \Big\}\ \ .
\nonumber
\end{eqnarray}

\begin{figure}[tb]
\psfig{figure=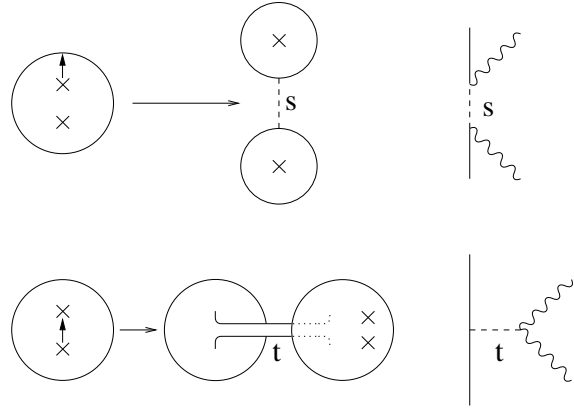,width=\hsize}
\caption{Factorization of the closed string two-point
function.}
\label{figb}
\end{figure}

The amplitude found found in (\ref{beta}) exhibits the Regge-pole
structure familiar from the Veneziano amplitudes.  The poles in the
$s$-channel occur as one of the vertex operators approaches the
boundary of the world sheet, while the $t$-channel poles occur as the
vertex operators approach each other.  (See figure \ref{figb}). The
fact that the amplitude can be expanded {\em either} in the
$s$-channel or in the $t$-channel poles is a characteristic feature of
world-sheet duality.  Another consequence of this structure is the
behavior of the amplitude for large energies at fixed scattering
angle.  There, one can apply the Stirling's approximation
$$\Gamma(u) = \sqrt{2 \pi} u^{u-1/2}e^{-u}$$
and express the scattering amplitude as
$$
A = (s a_1  - t a_2) \exp\left(-\frac{\alpha'}{2} s F(\phi) \right)
$$
where
$$F(\phi) = 
 \sin^2 \frac{\phi}{2} \log \sin^2\frac{\phi}{2}
+\cos^2 \frac{\phi}{2} \log \cos^2\frac{\phi}{2}.
$$
We have re-instated the factor of $\alpha'$.  The form factor
falls exponentially at a rate proportional to $\alpha'$, 
implying that the effective thickness of D-branes is of order the
string length and increases with the energy.
This is to be contrasted with the scattering off a point
particle in quantum field
theory, where the  fixed angle amplitude falls off as a power of $s$.

Scattering amplitudes with the Regge-pole structure and the
exponential fall-off of the form factors is a general feature of all
the D-branes except for the D-instanton (sometimes referred to as the
$-1$ brane).  The D-instantons are described by world sheets with
Dirichlet boundary condition imposed on {\em all} the coordinates
$X^M$, $0 \le M \le 9$.  The fact the they are special can be
understood on kinematical grounds.  A D-instanton breaks translation
invariance in all directions including the time.  Since there are no
`parallel' directions, one of the kinematic invariant variables,
$$ s = p_{1 \parallel}^2= p_{2 \parallel}^2\ ,$$
is meaningless and should be set to zero.  In this limit,
$$\frac{\Gamma(s) \Gamma(t)}{\Gamma(1+t+s)} \rightarrow \frac{1}{st},$$
and the total amplitude simplifies to 
\begin{eqnarray}
A &=& - \frac{1}{t}\tr(\varepsilon_1) p_1 \inn \varepsilon_2 \inn p_1
- \frac{1}{t}\tr(\varepsilon_2) p_2 \inn \varepsilon_1 \inn p_2 \nonumber \\
&& + \frac{1}{t} p_1 \inn (\varepsilon_2 - \varepsilon_2^T)
\inn (\varepsilon_1 - \varepsilon_1^T) p_2 \nonumber \\
&&-\frac{s+t}{s} \tr(\varepsilon_1) \tr(\varepsilon_2) \nonumber \\
&&- \frac{1}{2} (\varepsilon_1 - \varepsilon_1^T)  \inn 
(\varepsilon_2 - \varepsilon_2^T)
\end{eqnarray}
The amplitude no longer exhibits Regge-pole structure and resembles a
field theory amplitude instead.  We will later show that the
D-instanton acts as a source for the dilaton and the R-R scalar, but
leaves the Einstein metric flat. The dilaton two-point function
diverges. Presumably, this divergence is cancelled against the
divergence in the 
disconnected diagram where one of the operators
is inserted on an annulus, and the other on a disk. 
The divergence of the disconnected diagram occurs in the limit
where the inner boundary of the annulus 
shrinks to a point, and we expect a cancellation in the spirit of
the Fischler-Susskind mechanism.
\cite{Fischler:1986}. The anti-symmetric tensor field amplitude is
\begin{figure}
\centerline{\psfig{figure=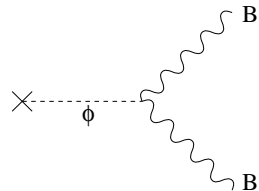,width=1.3in}}
\caption{Scattering of the Kalb-Ramond field by a D-instanton.}
\label{figd}
\end{figure}
$$A = \frac{1}{t} p_1 \inn B_2 \inn B_1 \inn p_2 - 
\frac{1}{2} \tr (B_1 \inn B_2).$$
This is precisely how the NS-NS antisymmetric tensor field will
scatter off the dilaton background created by the
D-instanton, due to the coupling
$${\cal L} = \phi H_{\mu \nu \lambda} H^{\mu \nu \lambda}$$
in the supergravity action (see figure \ref{figd}).  The graviton
scattering amplitude vanishes.  This is consistent with the fact that
there is no tree-level vertex coupling two on-shell gravitons to a
dilaton in the supergravity Lagrangian.

\section{Elastic scattering of massless R-R states off D-branes}
\label{rr}

In this section we extend the analysis of closed string scattering to
the gauge bosons from the R-R sector \cite{GHKM}.  The two-point
amplitude is of the general form (\ref{corr}) encountered in the
previous section.

In the canonical ghost picture, the vertex operator of an R-R gauge
boson with $m$-form field strength polarization $F_{(m)}$ and momentum
$k$ is given by
$$
\begin{array}{ll}
V(z,\bar{z}) = 
 \fslash_{(m)}^{\ \alpha}{}_\beta 
 V_{(-1/2)\, \alpha }(z)
 \bar{V}_{(-1/2)}^{\beta}(\bar{z})
 & {\rm (IIA)} \\
V(z,\bar{z}) =
 \fslash_{(m)}^{\alpha\beta}
 V_{(-1/2)\, \alpha }(z)
 \bar{V}_{(-1/2)\,\beta}(\bar{z})
   & {\rm (IIB)}
\end{array}                                              
$$
with
\begin{eqnarray}
V_{(-1/2)\, \alpha}(z) & = & 
e^{-\phi(z)/2} S_\alpha(z) e^{i k X(z)} \nonumber \\
V_{(-1/2)}^\beta(z) & = & e^{-\phi(z)/2} S^\beta(z) e^{i k X(z)} \nonumber
\end{eqnarray}
and
$$
\fslash_{(m)} = \frac{1}{m!} F_{\mu_1 \ldots \mu_m} 
   \gamma^{\mu_1} \cdots \gamma^{\mu_m} \ .           
$$
We have distinguished between the type IIA theory where the allowed
values of $m$ are 2 and 4, and the type IIB theory where
$m=1,3,5$.
Note that the two-point function of these
vertex operators on a disk has precisely the right amount of
total superghost charge.

We will use a representation in which the $32 \times 32$ Dirac gamma
matrices are off-diagonal:
$$
\gamma^\mu = \left( \begin{array}{cc}
                           0 & \gamma^{\mu\alpha\beta} \\
                           \gamma^\mu_{\alpha\beta} & 0
                    \end{array} \right) \ .
$$
We normalize the $\gamma^\mu$ so that 
$\{\gamma^\mu,\gamma^\nu\} = -2 \eta^{\mu\nu}$, and we pick our
representation so that
$$
\gamma_{11} = \gamma^0 \cdots \gamma^9 = 
              \left( \begin{array}{cc}
                              1 &  0 \\
                              0 & -1 
                     \end{array} \right) \ .
$$

The main new ingredient in the discussion of the R-R sector are
the spin fields, which assume a simple form
upon bosonization of the world-sheet fermions,
$$
S^{\alpha}(z) = C^\alpha e^{i\lambda^{\alpha}_i \phi_i(z)}$$
$$\frac{i}{\sqrt{2}}\left( \psi^{2i-1}(z) \pm i \psi^{2i}(z) \right) = e^{\pm i \phi_i(z)}$$
where $i = 1 \ldots 5$, and $\lambda^\alpha$'s are the weight vectors in
the chiral or anti-chiral conjugacy class of $SO(1,9)$. 
The cocycle operators, $C^\alpha$, impose the correct anticommutation
relations. The Dirichlet or Neumann boundary
conditions on $\psi$ translate 
into the following boundary conditions on the spin
fields, 
$$
\begin{array}{ll}
\bar{S}^\alpha(\bar{z}) = M^{\alpha\beta} S_\beta(z) 
& {\rm (type\ IIA)}  \\
\bar{S}_\alpha(\bar{z}) = M_\alpha{}^\beta S_\beta(z) 
& {\rm (type\ IIB)} \ .
\end{array}                                          
$$
where
$$M = \gamma^0 \gamma^1 \ldots \gamma^p.$$
The appearance of such a boundary condition was almost inevitable
given the $SO(1,p) \times SO(9-p)$ symmetry of the theory.  Just as
we did in the previous section, we can extend the definition of spin
fields from ${\cal H}^+$ to the entire complex plane 
by defining
$$
\begin{array}{ll}
S_{\alpha}(z) = \left\{
\begin{array}{ll}
S_{\alpha}(z) & z \in  {\cal H}^+ \\
M_{\alpha \beta} \bar{S}^\beta(z) & z \in {\cal H}^-
\end{array}
\right. & {\rm (type\ IIA)}  \\
S_{\alpha}(z) = \left\{
\begin{array}{ll}
S_{\alpha}(z) & z \in  {\cal H}^+ \\
{M_{\alpha}}^{\beta} \bar{S}_\beta(z) & z \in {\cal H}^-
\end{array}
\right. & {\rm (type\ IIB)} 
\end{array}
$$
In terms of the world sheet 
variables extended to the entire complex plane, the
amplitude becomes
\begin{eqnarray}
\lefteqn{A = \int \frac{dz_1 d\bar{z}_1 dz_2 d\bar{z}_2}{V_{CKG}}
(M \fslash^{(1)})^{\alpha \beta}
(M \fslash^{(2)})^{\gamma \delta}} \nonumber \\
&& \langle
V_{(-1/2)\alpha}(p_1,z_1)
V_{(-1/2)\beta}(p_1,\bar{z}_1) \nonumber \\
&& \qquad\qquad
V_{(-1/2)\gamma}(p_2,z_2)
V_{(-1/2)\delta}(p_2,\bar{z}_2)
 \rangle \nonumber 
\end{eqnarray}
The four-point
amplitude of the spin fields, which is needed here,
was calculated in \cite{FMS,KLLSW87},
\begin{eqnarray}
\lefteqn{
\langle S_\alpha(z_1) S_\beta(z_2) 
S_\gamma(z_3) S_\delta(z_4)\rangle = } \nonumber \\
&&  \frac{z_{14} z_{23} \gamma^\mu_{\alpha\beta} \gamma_{\mu\gamma\delta} - 
      z_{12} z_{34} \gamma^\mu_{\alpha\delta} \gamma_{\mu\beta\gamma} }{
      2 (z_{12} z_{13} z_{14} z_{23} z_{24} z_{34})^{3/4} }.   
\label{fourS}
\end{eqnarray}
The rest of the calculation follows exactly the same steps as
in the previous section, and we find that the amplitude has an
analogous form,
\begin{equation}
A = \frac{\Gamma(s) \Gamma(t)}{\Gamma(1+s+t)}\left[(s+t)P_1+s P_2\right]
\label{rramplitude}
\end{equation}
where $P_1$ and $P_2$ arise from two terms in (\ref{fourS}),
\begin{eqnarray}
P_1 &=& 
    = \tr \left( P \fslash_{(m)}^{(1)} M \gamma^\mu \right) 
      \tr \left( P \fslash_{(n)}^{(2)} M \gamma_\mu \right)  
          \nonumber \\
P_2 &=& 
    = \tr \left( P \fslash_{(m)}^{(1)} M \gamma^\mu 
             \fslash_{(n)}^{(2)} M \gamma_\mu \right) \ . \nonumber
\end{eqnarray}
and $P$ is the chiral projection operator
$$P = \frac{1 + \gamma_{11}}{2}$$

The final expression (\ref{rramplitude}) for the scattering of a
massless R-R boson has the same Regge-pole structure as the amplitude
of massless NS-NS bosons, (\ref{beta}).  Since the R-R bosons couple
to the R-R charge, we conclude that the effective size of the D-brane
R-R charge distribution is of the order $\sqrt{\alpha'}$, when
measured by the scattering of massless closed strings.

\section{Scattering off black $p$-branes in supergravity}
\label{pbrane}

In this
section we compare the scattering off D-branes
with the scattering off $p$-brane solitons of low-energy supergravity.
The fact
that two amplitudes agree for large impact parameters provides a
dynamical evidence that the D-branes are indeed the string theoretic
realizations of the supergravity $p$-branes.

Let us begin by reviewing the basic properties of the extreme R-R charged
$p$-brane solutions of the supergravity equations of motion.  General
solutions of this type were first written down in \cite{hs}:
\begin{eqnarray}
d s^2 &=& A^{-1/2} \left( -d t^2 + d x_1^2 + \ldots + d x_p^2 \right) 
\nonumber \\
&&\qquad +  A^{1/2} \left( d y^2 + y^2 d \Omega_{8-p}^2 \right)  \nonumber \\
e^{-2 \phi} &=& A^{(p-3)/2}  \nonumber \\
F_{(p+2)} &=& \frac{Q}{y^{8-p}} A^{-2} d t \wedge d x_1 \wedge \ldots 
   \wedge d x_p \wedge d y \nonumber  \\
\label{SolutionP}
\end{eqnarray}
where $F_{(p+2)}$ is the R-R field strength coupling to the brane, and 
\begin{equation}
A = 1 + \frac{2}{7-p} \frac{Q}{y^{7-p}} \ .                  
\label{DefineA}
\end{equation}
While \cite{hs} discussed only $0\leq p\leq 6$, solutions
(\ref{SolutionP}) and (\ref{DefineA}) may be extrapolated in an
obvious way to $p=7$ \cite{ggp} and $p=8$.  For the $7$-brane the
solution is (\ref{SolutionP}) with
$$
A = 1 + 2 Q\ln(y/y_0) \ ,                  
$$
while for the $8$-brane,
$$
A = 1 + 2 Qy= 1+ 2Q|x_9| \ .                  
$$
A new feature we find for $p=7$ and $8$ is that $A$ grows with the distance
from{} the $p$-brane. Thus, the geometry is not asymptotically flat.

The D-instanton ($p=-1$) is a special case because it requires a
Euclidean continuation \cite{GHKM,ggp}.  The D-instanton is a source
of the R-R scalar, which is essentially a ten-dimensional axion. In
order to make its axionic properties manifest, we are going to use its
dual, $8$-form, description. Our goal, therefore, is to find an
instanton solution to the following euclidean action,
$$
S= \int d^{10} x \sqrt G \biggl [e^{-2\phi} (-R+ 4 (\partial \phi)^2 )
+\frac{1}{2\cdot 9!} F_{(9)}^2 \biggr ]
$$
It is not hard to verify that the following is a solution:
\begin{eqnarray}
d s^2 &=&  A^{1/2} \left( d y^2 + y^2 d \Omega_9^2 \right)  \nonumber \\
e^{-2 \phi} &=& A^{-2}  \nonumber \\
F_{(9)} &=& \frac{Q}{y^9} A^{-2} * d y \nonumber
\end{eqnarray}
where 
$$
A = 1 + \frac{1}{4} \frac{Q}{y^8} \ . 
$$
This solution has the following interesting feature: the Einstein
metric,
$$
g^{\mu \nu} = G^{\mu \nu} e^{-\phi/2} = \delta_{\mu \nu}\ ,
$$
is flat!  Since the Einstein metric describes the physical
gravitational field, we conclude that the R-R charged instanton in 10
dimensions emits a dilaton, but no gravitational field. This explains
the vanishing of the graviton scattering amplitude off of a
D-instanton.

In order to compute the scattering at large impact parameter, we
expand the background in powers of
\begin{eqnarray}
\lambda &=& \frac{1}{7-p} \frac{Q}{y^{7-p}} \nonumber \\
&=& 
\frac{2 \pi^{(9-p)/2} }{ \Gamma((9-p)/2) }
  \int \frac{d^{9-p} q}{(2 \pi)^{9-p}} \, e^{i q \cdot y} \frac{Q}{q^2} \ .
\nonumber
\end{eqnarray}
To first order in $\lambda$, the string metric is given by
$$G^{\mu \nu} = \eta^{\mu \nu} - \lambda D^{\mu \nu}$$ 
The large impact parameter limit of the amplitude for strings to scatter
off a soliton background is obtained by expanding 
the relevant part of the action,
$$S = \int d^{10}x \sqrt{G} \frac{1}{2 \cdot n!} F_{(n)}^2$$ 
to leading order in $\lambda$.  To this order, we find
\begin{eqnarray}
\lefteqn{\delta S =
 \lambda \left\{ 
(\tr D) F^{(1)}_{\mu_1  \ldots \mu_n} F^{(2)\,\mu_1
\ldots \mu_n} \right.} \label{field}
 \\
\lefteqn{ + \left. n D_{\mu_1 \nu_1} \eta_{\mu_2 \nu_2} 
\ldots \eta_{\mu_n \nu_n} 
F^{(1)\,\mu_1  \ldots \mu_n}
F^{(2)\,\nu_1  \ldots \nu_n}
\right\}}  \nonumber
\end{eqnarray}
To make explicit comparisons with string theory we expand this
expression in terms of $SO(1,p)\times SO(9-p)$ invariants,
\begin{eqnarray}
\lefteqn{\delta S \sim}\nonumber \\
\lefteqn{\quad {1\over t}
\sum_{a+b=n}c^{(a,b)} F^{(1)}_{m_1  \ldots m_a M_1 \ldots M_b}
F^{(2)\, m_1  \ldots m_a M_1 \ldots M_b}} \nonumber 
\end{eqnarray}
After some combinatorics, one finds from (\ref{field}) that
\begin{equation}
c_{\rm field}^{(a,b)} =
 (4 - p +a-b) \binomial{n}{a}
\label{fieldcoeff}
\end{equation}
In order to compare this with the large impact 
parameter limit of the string
theory amplitude, we examine the leading $t$-channel pole 
in equation (\ref{rramplitude}),
$$A = \frac{1}{t}(P_1 + P_2).$$
In order to express this in terms of $SO(1,p) \times SO(9-p)$ invariants
as we did for the field theory amplitudes, we must evaluate the
trace of the $\gamma$-matrices.  
The main formula one uses is the general
(anti)-commutator of anti-symmetrized gamma matrices:
\begin{eqnarray}
\lefteqn{\left[ \gamma^{[\mu_1} \cdots \gamma^{\mu_m]},
       \gamma_{[\nu_1} \cdots \gamma_{\nu_n]} \right]_{(-1)^{mn+1}} =}
  \nonumber \\
 & & \sum_{j=1}^m (-1)^{1+mj+j(j+1)/2} \binomial{m}{j} \binomial{n}{j} 
\nonumber \\
 &&
      2^j j! \, \delta^{[\mu_1}_{[\nu_1} \cdots \delta^{\mu_j}_{\nu_j} 
      \gamma^{\mu_{j+1}} \cdots \gamma^{\mu_m]} 
      \gamma_{\nu_{j+1}} \cdots \gamma_{\nu_n]} .\nonumber
\end{eqnarray}
Using this identity, one can show that
\begin{eqnarray}
\lefteqn{P_1+P_2 =} \nonumber \\
&&-32\, c_{\rm string}^{(a,b)} \,F^{(1)}_{m_1  \ldots m_a M_1 \ldots M_b}
F^{(2)\, m_1  \ldots m_a M_1 \ldots M_b} \nonumber
\end{eqnarray}
with
\begin{eqnarray}
c^{(a,b)}_{\rm string} &=&
8 \left(\rule{0ex}{3ex} (p+2)\, \delta_{n-p-2}\, \delta_{b-1}  
+ \delta_{n-p} \,\delta_{b} \right. \nonumber \\
&& -\delta_{n+p-8} \,\delta_{a} -(10-p)  \,\delta_{n+p-10}  
\left.\rule{0ex}{3ex} \delta_{a-1} \right) \nonumber \\
&&- (-1)^{np+p(p+1)/2+n(n+1)/2+a}  \cdot \nonumber \\
&&  \qquad (4-p+a-b) \binomial{n}{a}.
\nonumber
\end{eqnarray}
It is a somewhat non-trivial fact that $c^{(a,b)}_{\rm field}$ and
$c^{(a,b)}_{\rm string}$ are identical! Altogether there
are 25 different scattering processes where our field theory
analysis serves as a check on the string theory computation: $n=1,3,5$
for $p$ odd and $n=2,4$ for $p$ even, with $-1 \leq p \leq 8$.  We
have checked the agreement between string theory and field theory in
all of the $25$ possible cases. 
The consistent agreement in the large impact parameter scattering is
a strong piece of evidence for regarding the extreme black $p$-branes
of supergravity as the
low-energy descriptions of the Dirichlet $p$-branes.

\section{Type I four-point amplitudes and D-branes}
\label{openclosed}

An astute reader may have noticed that our closed string two-point
scattering amplitudes closely resemble the type I open string
four-point amplitudes.  This relation was made precise in a
very nice paper by Garousi and Myers \cite{gm}.  Upon appropriately
fixing the residual conformal Killing volume, they found that
the two amplitudes are {\em identical} upon a certain identification
between the momenta and polarizations.

Consider a vertex operator for the vector boson of type I theory with
momentum $k$ and polarization $\zeta$, which is to be integrated over
the real axis.
\begin{equation}
V_0 = \zeta_\mu (\frac{1}{2}\partial_t X^\mu +
i\, 2k\inn \psi \psi^\mu) e^{i k\cdot X} (\sigma)
\label{openvertex}
\end{equation}
Using the world sheet doubling trick, we may write the operators in
the $-1$ and $0$ picture in terms of the holomorphic only,
\begin{eqnarray}
V_{-1}^\mu(z,2k) &=& 
e^{-\phi} \psi^\mu e^{i\, 2k \cdot X} (z) \nonumber \\
V_0^\mu(z,2k) &=& (\partial X^\mu +
i\, 2k\inn\psi\, \psi^\mu) e^{i\, 2k \cdot X} (z) 
\nonumber
\end{eqnarray}
Let us consider the $2 \rightarrow 2$ scattering of type I open strings
with momentum $k_i$ and  polarizations $\zeta_i$ for $i=1\ldots4$.
The amplitude is written explicitly as
\begin{eqnarray}
\lefteqn{A(\zeta_1, k_1; \zeta_2, k_2; \zeta_3, k_3; \zeta_4, k_4)} \nonumber \\
&=& 
\int \frac{dx_1 dx_2 dx_3 dx_4}{V_{CKG}}
\nonumber \\
&& \qquad
\langle \zeta_1\inn V_{0}(2k_1,x_1)\,
\zeta_2\inn V_{0}(2k_2,x_2) \, \nonumber \\
&& \qquad\qquad
\zeta_3\inn V_{-1}(2k_3,x_3) \,
\zeta_4\inn V_{-1}(2k_4,x_4) \rangle
\nonumber
\end{eqnarray}
For future use, it is convenient to introduce the Mandelstam variables
\begin{eqnarray}
s &=& 4k_1 \inn k_2 = 4 k_3 \inn k_4, \nonumber \\ 
t &=& 4k_1 \inn k_4 = 4 k_2 \inn k_3, \nonumber \\
u &=& 4k_1 \inn k_3 = 4 k_2 \inn k_4. \nonumber
\end{eqnarray}
The traditional method for cancelling the conformal 
Killing volume is to set the vertex operators at 
$$ \{x_1,x_2, x_3, x_4\} = \{0,x,1,\infty\}$$
and to integrate $x$ from 0 to 1. An alternative is to
fix the vertex operators at
$$ \{x_1,x_2, x_3, x_4\} = \{-x,x,1,-1\}$$
The four-point amplitude then takes the form
\begin{eqnarray}
\lefteqn{A = \int_0^1 \, dx \, (1-x^2)
\zeta_{1\mu} \zeta_{2\nu} \zeta_{3\sigma} \zeta_{4\kappa}}\nonumber \\
\lefteqn{\langle
V_{-1}^\mu (k_1,-x)
V_{-1}^\nu (k_2,x)
V_{0}^\sigma (k_3,1)
V_{0}^\kappa (k_4,-1)
 \rangle} \nonumber \\ 
\label{openfour}
\end{eqnarray}
This is identical to the integral encountered in (\ref{2camp}) under
the identification
\begin{equation}
\begin{array}{ll}
2k_1^\mu\rightarrow (D \inn p_1)^\mu &
2k_2^\mu\rightarrow p_1^\mu \\
2k_3^\mu\rightarrow p_2^\mu &
2k_4^\mu\rightarrow (D\inn p_2)^\mu \\
\multicolumn{2}{l}
{\zeta_{2\mu}\,\otimes\,\zeta_{1\nu} \rightarrow \varepsilon_{1\mu\lambda}D^\lambda{}_\nu} \\
 \multicolumn{2}{l}{
\zeta_{3\mu}\,\otimes\,\zeta_{4\nu} \rightarrow \varepsilon_{2\mu\lambda}D^\lambda{}_\nu.}
\end{array}
\label{ident}
\end{equation}
The result of evaluating (\ref{2camp}) was
$$A = \frac{\Gamma(s)\Gamma(t)}{\Gamma(1+s+t)} (s a_1 - t a_2)$$
On the other hand, the
type I four-point function has the well known form \cite{SchwarzPhysRep}
\begin{eqnarray}
A &=& \frac{\Gamma(s)\Gamma(t)}{\Gamma(1+s+t)} K(\zeta_1,k_1;\zeta_2,k_2;\zeta_3,k_3;\zeta_4,k_4) \nonumber \\
\label{openns4}
\end{eqnarray}
where $K(\zeta_1,k_1;\zeta_2,k_2;\zeta_3,k_3;\zeta_4,k_4)$ is the  standard
kinematic factor,
\begin{eqnarray}
\lefteqn{K(\zeta_1,k_1;\zeta_2,k_2;\zeta_3,k_3;\zeta_4,k_4) =}\nonumber \\
\lefteqn{\ \ -16k_2\inn k_3\,k_2\inn k_4\ \zeta_1\inn\zeta_2\,\zeta_3\inn\zeta_4}
\nonumber\\
\lefteqn{\qquad  -4k_1\inn k_2\,
(\zeta_1\inn k_4\,\zeta_3\inn k_2\,\zeta_2\inn\zeta_4
+\zeta_2\inn k_3\,\zeta_4\inn k_1 \,\zeta_1\inn\zeta_3} \nonumber \\
\lefteqn{ \qquad\qquad +
 \zeta_1\inn k_3\,\zeta_4\inn  k_2\,\zeta_2\inn\zeta_3
 +\zeta_2\inn k_4\,\zeta_3\inn k_1\,\zeta_1\inn\zeta_4)}
\nonumber\\
&&\qquad +\Big\{1,2,3,4\rightarrow 1,3,2,4\Big\} \nonumber \\
&&\qquad +\Big\{1,2,3,4\rightarrow 1,4,3,2\Big\}.
\label{kinfactor}
\end{eqnarray}
The fact that the amplitude for scattering off a D-brane (\ref{2camp})
and the open string four point amplitude (\ref{openfour}) take on
identical forms allows us to identify the 
kinematic factors,
\begin{equation}
(s a_1 - t a_2) = K(\zeta_1,k_1;\zeta_2,k_2;\zeta_3,k_3;\zeta_4,k_4)
\label{obscure}
\end{equation}
One can verify that $a_1$ and $a_2$ found in section \ref{nsns} indeed
satisfy (\ref{obscure}) under the identification (\ref{ident}).

Now, the open string four-point amplitudes have been computed for
all
combinations of NS and Ramond external states.
The amplitudes have the general structure 
(\ref{openns4}) 
while the kinematic factors are given by
\cite{SchwarzPhysRep}:
\begin{eqnarray}
\lefteqn{K_2(u_1,\,u_2,\,u_3,\,u_4)=
-2\,k_1\inn k_2\,\bar{u}_2 \gamma^\mu u_3\,\bar{u}_1\gamma_\mu u_4 }
\nonumber \\
&&+2\,k_1\inn k_4\,\bar{u}_1\gamma^\mu u_2\,\bar{u}_4\gamma_\mu u_3 \label{kin2}
\\
\lefteqn{K_3(u_1,\,\zeta_2,\,\zeta_3,\,u_4)=}\nonumber \\
&&2i\sqrt{2}\,k_1\inn k_4\,\bar{u}_1\gamma\inn \zeta_2\gamma\inn(k_3+k_4)\gamma\inn\zeta_3 u_4
\nonumber \\
&&-4i\sqrt{2}\,k_1\inn k_2\,\left(\bar{u}_1\gamma\inn\zeta_3u_4\,k_3\inn\zeta_2
\right.\nonumber \\
&&-\bar{u}_1\left.\gamma\inn\zeta_2u_4\,k_2\inn\zeta_3-\bar{u}_1\gamma\inn k_3u_4\,\zeta_2\inn\zeta_3\right) \label{kin3} \\
\lefteqn{K_4(u_1,\,\zeta_2,\,u_3,\,\zeta_4)=}\nonumber \\
&&-2i\sqrt{2}\,k_1\inn k_4\,\bar{u}_1\gamma\inn \zeta_2\gamma\inn(k_3+k_4)\gamma\inn\zeta_4 u_3\nonumber \\
&&-2i\sqrt{2}\,k_1\inn k_2\,\bar{u}_1\gamma\inn
\zeta_4\gamma\inn(k_2+k_3)\gamma\inn\zeta_2 u_3 \label{kin4}
\ \ .
\end{eqnarray}
The observation that the two point amplitude of a closed string in a
background of a D-brane is identical to the four point function of type
I theory is extremely useful because one can apply this identification
to compute scattering amplitudes of all types of closed strings:
NS-NS, R-R, NS-R and R-NS. 
For the sake of illustration, consider a process where
a massless NS-NS boson with momentum $p_1$ and polarization
$\varepsilon_{\mu \nu}$ is absorbed and a R-R boson with momentum
$p_2$ and polarization $F$ is emitted by the D-brane.  The amplitude
for such a process is given by
$$A = 
\frac{\Gamma(s) \Gamma(t)}{\Gamma(1+s+t)}
K_3(u_1,\,\zeta_2,\,\zeta_3,\,u_4) 
$$
where we identify
$$
\begin{array}{ll}
2k_1^\mu\rightarrow p_1^\mu &
2k_2^\mu\rightarrow p_2^\mu \\
2k_3^\mu\rightarrow (D\inn p_2)^\mu &
2k_4^\mu\rightarrow (D\inn p_1)^\mu \\
\multicolumn{2}{l}
{u_{1A}\,\otimes\,u_{4B} \rightarrow (P\,\fslash M)_{AB}} \\
\multicolumn{2}{l}
{\zeta_{2\mu}\,\otimes\,\zeta_{3\nu}\hphantom{B}
\rightarrow \varepsilon_{2\mu\lambda}D^\lambda{}_\nu\ \ } 
\end{array}
$$
Using similar identifications, Garousi and Meyers were able to derive
all $1\rightarrow 1$ amplitudes for closed strings.

Let us now consider the process illustrated in figure \ref{figa}.B.
This is the four-point function of open strings with ends attached to
the D-brane. The vertex operator for such an open string is obtained
by T-dualizing the vertex operator (\ref{openvertex}) along the
directions transverse to the brane,
$$
V_0^M = (\frac{1}{2}\partial_n X^M + i\, 2 k\inn \psi \psi^M) e^{i k\cdot X}
(\sigma).
$$
Now, since the boundary conditions along these directions are Dirichlet,
$$
\begin{array}{lll}
X^M(z) & = & -\bar{X}^M(\bar{z}) \\
\psi^M(z) & = & -\bar{\psi}^M(\bar{z}) \\
\end{array} \qquad (z,\bar{z}) \in \partial {\cal H}^+
$$
the vertex operator, when expressed in strictly holomorphic variables,
becomes identical to what we found in the case of Neumann boundary
condition (with an extra constraint that the momentum $k$ 
be parallel to the brane.)  
\begin{eqnarray}
V_{-1}^M(z,2k) &=& 
e^{-\phi} \psi^M e^{i\, 2k \cdot X} (z) \nonumber \\
V_0^\mu(z,2k) &=& (\partial X^M +
i\, 2k\inn\psi\, \psi^M) e^{i\, 2k \cdot X} (z) 
\nonumber
\end{eqnarray}
Thus, the amplitude of figure \ref{figa}.B is
identical to the open string
four-point function,
$$A = \frac{\Gamma(s)\Gamma(t)}{\Gamma(1+s+t)} K$$
with the restriction that all the momenta 
are parallel to the brane.  
This is hardly a surprise given the fact that
the low-energy effective dynamics on a Dirichlet $p$-brane is 
described by a
dimensional reduction of $N=1$ supersymmetric
Yang-Mills theory from 10-dimensions to $p+1$
dimensions.

\section{Absorption and Hawking emission by D-branes}
\label{decay}

Finally, let us consider the process illustrated in figure
\ref{figa}.C, that is, the absorption of a massless closed string
state leading to a pair creation of open strings attached to the
D-brane \cite{decay}.  To begin, let us focus on the NS sector and
assign momenta $p_1$ and $p_2$ to the open string states and momentum
$q$ to the closed string. 
The open string momenta $p_1$ and $p_2$ are 
restricted to lie within the D-brane world volume.
The D-brane kinematics is such that only
the longitudinal momentum is conserved,
\begin{equation}
p_1 + p_2 + q_\parallel = 0 \ .
\end{equation}
Ordinarily in massless three-point amplitudes conservation of momentum
constrains the kinematics completely but, in the presence of D-branes,
the non-conservation of momentum in the directions transverse to the brane
leaves some freedom.  Here, from{} the conservation of the longitudinal
momentum, it follows that there is exactly one kinematic invariant in
this problem, which we call $t$,
\begin{equation}
t = 2 p_1 \inn q = 2 p_2 \inn q = - 2 p_1 \inn p_2 \ .
\end{equation}

The leading order contribution to this amplitude is evaluated on a
disk with two operators on the boundary and one in the bulk.  We
proceed by mapping the disk to the upper half-plane, 
\begin{equation}
A = \int \frac{d z_1\, d z_2\, d^2 z_3}{V_{CKG}} \langle 
V_1(z_1)
V_2(z_2)
V_3(z_3,\bar{z}_3)
\rangle \label{gen.amp}
\end{equation}
where $z_1$ and $z_2$ are integrated only along the real axis.  The
vertex operators are the same as the ones used in the preceding sections
and, after the doubling of the world sheet, the amplitude becomes
\begin{eqnarray}
\lefteqn{A = \int \frac{dz_1\, dz_2\, dz_3\, d\bar{z}_3}{V_{CKG}}\,
\xi_\mu^1 \xi_\nu^2 \varepsilon_{\sigma \lambda} {D^\lambda}_\eta}  \\
\lefteqn{\ \langle V_0^\mu(z_1,2p_1) 
        V_0^\nu(z_2,2p_2) 
        V_{-1}^\sigma(z_3,q) 
        V_{-1}^\eta(\bar{z}_3, D\inn q) \rangle} \nonumber 
\label{integral.expression}
\end{eqnarray}
This is precisely the form of the correlation function encountered
previously in the computation of the open string 4-point function
(\ref{openfour}), if we identify
\begin{equation}
\begin{array}{ll}
2 k_1    \rightarrow 2 p_1 &
2 k_2    \rightarrow 2 p_2  \\
2 k_3    \rightarrow q& 
2 k_4    \rightarrow D \inn q \\
\zeta_1  \rightarrow \xi_1  &
\zeta_2  \rightarrow \xi_2 \\
\multicolumn{2}{l}{\zeta_3 \otimes \zeta_4  \rightarrow \varepsilon \inn  D}  \\
\end{array}.
\label{nsns.identify}
\end{equation}
This allows us to use a short-cut similar to that found in
\cite{gm} for the
closed string two-point function. 
We fix the vertex operators at
$\{z_1, z_2, z_3, {\bar z_3}\} = \{-x, x, i,-i\}$, which corresponds to
placing the closed string vertex at $z=i$ and constraining the
open string vertex operators to lie symmetrically on the real axis.
Notice that this is related to the calculations
in section 1 by a change of variables, $y
=-ix$.  Recalling that there is only one kinematic invariant in
this problem, we set $s = 4 k_1 \inn k_2 = -2 t$.  After these
replacements the amplitude takes the from{}:
\begin{equation}
A=\int_{-\infty}^{\infty} dx \left[ \frac{(1 + x^2)^2}{16 x^2} \right]^t
\frac{1}{1+x^2} \left(a_1 + \frac{a_2}{2}\right)
\end{equation}
Performing the integral
we find 
\begin{equation}
A =  (-2 t  a_1 - t a_2)  \frac{\Gamma(-2t)}{\Gamma(1-t)^2}
\end{equation}
It is clear that $(-2t a_1 - t a_2)$ is the same kinematic factor $K =
(s a_1 - t a_2)$ that we encountered above, with $s = -2 t$. Thus, 
\begin{equation}
A =  \frac{\Gamma(-2t)}{\Gamma(1-t)^2}
 K(1, 2, 3)
 \label{main.result}
\end{equation}
where the kinematic factor is obtained from{} that of type I theory,
(\ref{kinfactor}), by the identifications (\ref{nsns.identify}).  This
is the main result of this section.

The explicit formula for the
Neveu-Schwarz amplitudes is, therefore,
\begin{eqnarray}
\lefteqn{K(1, 2, 3)  =}\nonumber \\
&&  \left[\rule{0ex}{2ex} t\,\left( - q\inn \xi_2\,\xi_1\inn  D \inn q\,
         g^{\mu \nu}   - 
      q\inn \xi_1\,\xi_2\inn  D\inn q\,g^{\mu\nu} \right. \right. \nonumber \\
&& -4\,\xi_1\inn \xi_2\,p_1^\nu \,p_2^\mu  - 
      4\,\xi_1 \inn \xi_2\,p_1^\mu \,p_2^\nu  \nonumber \\  
&&- 
      2\,p_1 \inn \xi_2\,q^\nu \,\xi_1^\mu  + 
      4\,q\inn \xi_2\,p_1^\nu \,\xi_1^\mu \nonumber \\
&& - 
      2\,p_1\inn \xi_2\, (D \inn q)^\mu \,\xi_1^\nu + 
      4\,\xi_2\inn  D\inn q\,p_1^\mu \,\xi_1^\nu  \nonumber \\
&& - 
      2\,p_2\inn \xi_1\,q^\nu \,\xi_2^\mu  + 
      4\,q\inn \xi_1\,p_2^\nu \,\xi_2^\mu  \nonumber \\
&& - \left.
      2\,p_2\inn \xi_1\,(D \inn q)^\mu \,\xi_2^\nu  + 
      4\,\xi_1 \inn  D \inn q\,p_2^\mu \,\xi_2^\nu  \right) \nonumber \\
&&\ \ \   +  \left.
   {t^2}\,\left( -\,\xi_1\inn \xi_2\,g^{\mu \nu} + 
      2\,\xi_1^\nu \,\xi_2^\mu  + 2\,\xi_1^\mu \,\xi_2^\nu
       \right) \rule{0ex}{2ex} \right] \nonumber \\
&& \qquad(\varepsilon \inn D)_{\mu \nu}
\end{eqnarray}
The gauge invariance follows automatically from{} 
that of the type I
kinematic factor.  When we take the polarization of the closed string
to be traceless and strictly transverse to the D-brane, the kinematic
factor simplifies considerably, and one is left with
\begin{equation}
A \sim \frac{\Gamma(-2t)}{\Gamma(1-t)^2} t^2
\left(\xi^1 \inn  \varepsilon  \inn \xi^2 +
\xi^2 \inn  \varepsilon  \inn \xi^1 \right)  
\end{equation}
In the low $t$ approximation, and for symmetric $\varepsilon$ (i.~e.
those describing gravitons), this amplitude reduces to 
$$A = t \varepsilon_{MN} \xi_1^M \xi_2^N$$
which coincides with the three point amplitude obtained from following
term in the effective action
$${\cal L} = \int d^{p+1}x \partial_m \phi^M 
\partial^m \phi^N G_{MN}$$
The fields $\phi^M (x^m)$ specify the transverse location of the D-brane.
This is the leading term in the static gauge expansion of
the Nambu-Goto action.

The amplitudes for states involving the R sector are again of the
general form found in (\ref{main.result}), with the kinematic factors
obtained from the type I ones, (\ref{kin2}-\ref{kin4}), under
appropriate identification of kinematic variables.  Let us illustrate
this with some examples.

First, consider pair creation of NS open strings
by an R-R closed string with $n$-form polarization
$F_{\mu_1 \mu_2 \ldots \mu_n}$.  In this case, the correlation
function is of the form
\begin{eqnarray}
\lefteqn{\xi_\mu \xi_\nu (P\fslash M)^{AB}} \nonumber \\
\lefteqn{\qquad\langle 
 V_0^\mu(z_1, 2p_1)\,
 V_{-1}^\nu(z_2, 2p_2)} \nonumber \\
\lefteqn{\qquad\qquad
 V_{-1/2\, A}(z_3,q)\, 
 V_{-1/2\, B}(\bar{z}_3,D \inn q)
\rangle} \nonumber
\end{eqnarray}
where just as in section \ref{rr},
\begin{eqnarray}
\fslash & =& \frac{1}{n!} 
F_{\mu_1 \mu_2 \cdots \mu_n} 
\gamma^{\mu_1} \gamma^{\mu_2} \cdots \gamma^{\mu_n}, \nonumber \\
M & = & \gamma^0 \gamma^1 \cdots \gamma^p, \nonumber \\
P &=& (1 + \gamma_{11})/2. \nonumber
\end{eqnarray}
The amplitude reduces to
\begin{equation}
A =  \frac{\Gamma(-2t)}{\Gamma(1-t)^2}K(1, 2, 3)\ , \label{genamp}
\end{equation}
with the kinematic factor obtained from{} that in
type I theory, (\ref{kin3}),
by the following substitutions,
$$
\begin{array}{ll}
2 k_1   \rightarrow D \inn q &
2 k_2   \rightarrow 2 p_1 \\
2 k_3   \rightarrow 2 p_2 &
2 k_4   \rightarrow q  \\
\zeta_2 \rightarrow \xi_1   &
\zeta_3 \rightarrow \xi_2  \\
\multicolumn{2}{l}{u_4 \otimes u_1 \rightarrow  P \fslash M} 
\ .\end{array}
$$
This gives
\begin{eqnarray}
\lefteqn{K(1, 2, 3)  = 
t\,
\tr\left[P \fslash M \right.}  \nonumber \\
&&\left.\left( \rule{0ex}{2ex} 
\gamma\inn \xi_1\gamma\inn(p_2+q/2)\gamma\inn\xi_2
+ 4( p_2\inn\xi_1 ) \gamma\inn\xi_2  \right. \right. \nonumber \\
&&\qquad \left. \left. \rule{0ex}{2ex} - 4(p_1\inn\xi_2) \gamma\inn\xi_1
- 4 (\xi_1\inn\xi_2) \gamma\inn p_2 \right) \right] \nonumber
\end{eqnarray}

Similarly, consider exciting two open string fermions with
polarizations $v_1^A$ and $v_2^B$ by an incident massless
NS-NS boson with
polarization $\varepsilon_{\mu \nu}$. Now, the correlation function
takes the form
\begin{eqnarray}
\lefteqn{v_1^A v_2^B (\varepsilon \inn D)_{\mu \nu}}\nonumber \\
&& \langle 
V_{-1/2\, A}(z_1, 2p_1)\, 
V_{-1/2\, B}(z_2, 2p_2)\,  \nonumber \\
&& \qquad V_0^\mu(z_3, q)\,
V_{-1}^\nu(\bar{z}_3, D \inn q)
\rangle \nonumber
\end{eqnarray}
The relevant kinematic factor is obtained from (\ref{kin3}) by
the following substitutions,
$$
\begin{array}{ll}
2 k_1  \rightarrow 2 p_2 &
2 k_2  \rightarrow q       \\ 
2 k_3  \rightarrow D \inn q &
2 k_4  \rightarrow 2 p_1     \\
u_4 \rightarrow  v_1 &
u_1 \rightarrow v_2  \\
\multicolumn{2}{l}
{\zeta_2  \otimes  \zeta_3 \rightarrow (\varepsilon \inn D)} \\
\end{array}
$$
Thus, the amplitude for two open string fermions to produce a NS-NS
closed string state is again of the general form (\ref{genamp}) with
the kinematic factor
\begin{eqnarray}
\lefteqn{K(1,2,3)= t\, (\varepsilon \inn D)_{\mu\nu}}\nonumber \\ 
&&\left[\rule{0ex}{2ex}
 \bar{v}_2\gamma^\mu \gamma\inn(D \inn q+ 2p_1)\gamma^\nu v_1
 + 4 \bar{v}_2\gamma^\nu v_1\,(D \inn q)^\mu  \right.\nonumber \\
&&\left. \rule{0ex}{2ex} - 4\bar{v}_2\gamma^\mu v_1\,q^\nu 
- 4\bar{v}_2(\gamma\inn D \inn q ) v_1\, g^{\mu \nu} 
\right]  \nonumber
\end{eqnarray}
It is straightforward to extend this program to other combinations of 
Neveu-Schwarz and Ramond vertex operators.

\begin{figure}[t]
\centerline{\psfig{figure=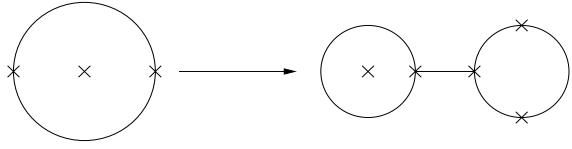,width=\hsize}}
\caption{Factorization of the world sheet giving rise to poles in
$t$.}
\label{fige}
\end{figure}

It is interesting that these amplitudes 
occupy an intermediate position between
the conventional three-point and four-point amplitudes 
as far as the number of
kinematic invariants goes. This is a consequence of the
`partial conservation of momentum' unique to D-branes.
These amplitudes have other interesting new features.  They
decay exponentially for large $t$, which
indicates the softness of
strings at high energies, 
and have a sequence of poles at half-odd-integer
values of $t$. What is surprising is that they also contain a sequence
of zeros for integer values of $t$.  The special role played by these
values of $t$ is related to the massive open string states
appearing in the operator product expansion of the open string vertex
operators when they collide on the world sheet (see figure
(\ref{fige})).
These states have masses
\begin{equation}
m^2 = n/\alpha' = n/2
\end{equation}
for integer $n$, and indeed
\begin{equation}
t = -2 p_1 \inn p_2 = -(p_1 + p_2)^2 = m^2 =  n/2.
\end{equation}
Interestingly, for even $n$ these amplitudes have zeros instead of
poles, indicating that these states do not propagate in the internal
line of figure 1.  Thus, due to the special kinematics of this
process, we find an interesting $Z_2$ selection rule. This interplay
between the zeros and the poles is intimately connected with the exponential
decay of the amplitude at large $t$.

\section{Conclusions}
\label{conclusion}

These lecture notes summarize some of the simplest perturbative calculations
that describe interactions of D-branes and elementary strings.  The results
reveal a consistent physical picture.  Since the D-branes are described
by open strings whose ends are attached to specified hyperplanes,
and couple to closed strings via the conventional open-closed string
interactions, they acquire the softness characteristic of strings.
Their effective thickness, observed with massless closed string
probes, is of order the string scale, $\sqrt{\alpha'}$, 
and grows with the energy.

One could wonder, however, whether there exists a smaller structure
underneath the `string halo' that we have described.  A number of
interesting recent works have reached the conclusion that the answer
is yes, and that this structure becomes apparent only when the
D-branes are probed with other D-branes whose velocity is very small
\cite{bachas,Lifschytz:1996a,%
Douglas:1996,Johnson:1996,DanFerrSund,PouliotKabat,DKPS96}.  This
effect seems to crucially depend on the cancellation of forces at rest
between certain types of D-branes, i.e., on the fact that we are
studying slow motion in a BPS saturated system.  The scattering of
D-branes from other D-branes is a complicated problem because, unlike
the questions considered here, it is entirely non-perturbative.  Its
study in the context of low energy field theory reveals a
non-perturbative length scale, $g^{1/3}\sqrt{\alpha'}$, which is
completely natural since it is the Planck scale of the 11-dimensional
M-theory.  We feel, however, that there is much that remains to be
learned about the structure of D-branes.  We need a clear
understanding of which scales govern various processes: at what stage,
for instance, do the shorter scales found in slow motion get smeared
by the string scale.  We should also keep in mind that, if the string
coupling $g$ turns out to be of order one, then the string scale and the
11-dimensional Planck scale turn out to be of the same order and should be
regarded as competing effects.  To conclude, the D-branes have
provided us with a remarkable new viewpoint on non-perturbative string
theory.  No doubt, much remains to be learned from this remarkable
tool.

\section*{Acknowledgements}

We are grateful to S. Gubser, J. Maldacena and L. Thorlacius
for collaborations on the material presented here.
This work was supported in part by the DOE grant
DE-FG02-91ER40671, the NSF Presidential Young Investigator Award
PHY-9157482, and the James S. McDonnell Foundation grant No. 91-48.

\end{document}